\documentclass[aps,prd,amsmath,amssymb,nofootinbib,nobibnotes,twocolumn]{revtex4-2}

\usepackage{graphicx}
\usepackage{dcolumn}
\usepackage{footmisc}
\usepackage[section]{placeins} 
\usepackage{bm}
\usepackage{xcolor}
\usepackage{float}
\usepackage{dblfloatfix} 
\usepackage{capt-of}    
\makeatletter
\renewcommand{\fnum@figure}{Figure \thefigure}
\makeatother

\usepackage{hyperref}
\usepackage{orcidlink}

\begin{document}

\setlength{\abovedisplayskip}{5pt}
\setlength{\belowdisplayskip}{5pt}
\setlength{\abovedisplayshortskip}{5pt}
\setlength{\belowdisplayshortskip}{5pt}
\setlength{\textfloatsep}{8pt}
\setlength{\floatsep}{6pt}
\setlength{\intextsep}{6pt}
\setlength{\abovecaptionskip}{4pt}
\setlength{\belowcaptionskip}{-4pt}

\preprint{}

\title{Three-Technology Ultra-High-Energy Neutrino Flavor Measurement}

\author{Alba Burgos-Mondéjar~\orcidlink{0009-0009-1439-928X}}
\email{ab2910@cam.ac.uk}
\affiliation{Harvard University, Department of Physics and Laboratory for Particle Physics and Cosmology, Cambridge, MA 02138, USA}

\author{William G. Thompson~\orcidlink{0000-0003-2988-7998}}
\email{will\_thompson@g.harvard.edu}
\affiliation{Harvard University, Department of Physics and Laboratory for Particle Physics and Cosmology, Cambridge, MA 02138, USA}

\author{Carlos A. Argüelles~\orcidlink{0000-0003-4186-4182}}
\email{carguelles@fas.harvard.edu}
\affiliation{Harvard University, Department of Physics and Laboratory for Particle Physics and Cosmology, Cambridge, MA 02138, USA}

\begin{abstract}
Ultra-high-energy (UHE) neutrinos, with energies above $100\,\mathrm{PeV}$, are powerful probes of fundamental physics, offering unique opportunities to test and constrain models beyond the Standard Model. A central challenge is achieving sensitivity to their flavor composition, which encodes information about production mechanisms at the source as well as potential new physics during propagation. We propose, for the first time, a three-technology strategy to measure the flavor composition of UHE neutrinos by leveraging the complementary strengths of planned and existing experiments. This approach integrates in-ice radio (ARA, ARIANNA, RNO-G, the IceCube-Gen2 radio array), Earth-skimming (TAMBO, POEMMA, Trinity, AUGER, GRAND200k), and in-ice optical Cherenkov detectors (the IceCube-Gen2 optical array).
We demonstrate the potential to achieve unprecedented sensitivity to UHE neutrino flavors. 
The method can be used to probe an energy-dependent neutrino flavor transition at the source across energies ranging from TeV to EeV. 
As such, this article highlights the discovery potential of a coordinated, multi-technology program for advancing both high-energy and ultra-high-energy neutrino astrophysics.
\end{abstract}

\maketitle

\section{Introduction}
\label{sec:intro}

The IceCube Neutrino Observatory has observed high-energy astrophysical neutrinos in the $\mathrm{TeV}$--$\,\mathrm{PeV}$ range ~\cite{IceCube:2013low}. Beyond this energy window, the long-anticipated ultra-high-energy (UHE) neutrinos~\cite{Berezinsky:1969erk}, with energies exceeding $100\,\mathrm{PeV}$, hold the promise of unveiling deeper insights into astrophysical phenomena. Despite decades of theoretical predictions, such neutrinos remain undetected by IceCube to this day~\cite{IceCube:2018fhm}.

In 2025, the KM3NeT collaboration reported the observation of an ultra-high-energy through-going muon, designated KM3-230213A, detected in 2023 by its ARCA detector. The reconstructed muon-neutrino energy of this event, $220^{+570}_{-110}\,\mathrm{PeV}$, along with its nearly horizontal trajectory, makes an atmospheric origin highly improbable and establishes it as the most energetic neutrino ever recorded~\cite{KM3NeT:2025npi}.
This remarkable detection is not without precedent, as it follows earlier UHE neutrino candidate events, including four sub-horizon events identified by ANITA-IV in 2016~\cite{ANITA:2020gmv}.  

The detection of such events suggests that progress in ultra-high-energy neutrino detection now hinges upon a new generation of large-scale telescopes that are transitioning from design to prototype stages.
Unlike IceCube, which detects only optical Cherenkov light, the next generation of detectors will employ three complementary techniques: in-ice radio techniques, optical Cherenkov detection, and Earth-skimming-neutrino detectors.

\textbf{In-ice radio} detectors search for nanosecond-scale radio pulses generated when UHE neutrino interactions in ice produce compact electromagnetic showers~\cite{Ackermann:2022rqc}. These showers emit coherent radio-frequency pulses (Askaryan radiation) at the Cherenkov angle, due to the excess negative charge in the cascade. The long attenuation length of radio waves in ice, on the order of kilometers, enables the detection of such signals over large volumes, making this technique sensitive to all three neutrino flavors. Pilot projects such as ARA~\cite{Allison:2011wk} and ARIANNA~\cite{Persichilli:2019wyh} have demonstrated the feasibility of such in-ice radio detectors, and the Radio Neutrino Observatory in Greenland (RNO-G) is currently under construction~\cite{RNO-G:2020rmc}. Looking ahead, an order-of-magnitude more sensitive radio array is being planned as part of IceCube-Gen2~\cite{IceCube-Gen2:2021rkf}.  

\textbf{Earth-skimming} arrays exploit the unique phenomenology of $\nu_\tau$ interactions. A UHE $\nu_\tau$ undergoing a charged-current interaction in mountains or Earth’s crust produces a tau lepton which, owing to its long lifetime at these energies, can travel tens to hundreds of kilometers before decaying. If the geometry is favorable, the tau emerges from the Earth and decays in the atmosphere, initiating an air shower. Even if the tau decays underground, $\nu_\tau$ regeneration ensures that another neutrino is produced, maintaining the possibility of detection. Several projects—TAMBO-5k (the nominal 5{,}000-detector configuration of TAMBO), POEMMA, Trinity (in its nominal 18-telescope configuration), AUGER and GRAND200k—are developing this strategy, offering a promising new channel for UHE neutrino astronomy~\cite{TAMBO:2025jio, POEMMA:2020ykm, Stepanoff:2025vys, GRAND:2018iaj, PierreAuger:2019ens}.

Finally, \textbf{in-ice optical Cherenkov} detection relies on Cherenkov radiation emitted by relativistic charged particles produced in neutrino interactions with nucleons and electrons. Charged-current interactions of electron and tau neutrinos, together with neutral-current interactions of all neutrino flavors, generate electromagnetic or hadronic showers, or short-lived tau leptons at low energies. The proposed extension of the IceCube observatory, IceCube-Gen2, will feature a new optical array covering eight cubic kilometers at depths between $1.4$ and $2.7\,\mathrm{km}$~\cite{Ackermann:2022rqc}. This upgrade will enhance the detector’s sensitivity to in-ice Cherenkov signals across TeV to EeV energies. It includes five detection channels: cascade-like events originating within the instrumented volume (cascades and double-cascades), as well as through-going track events (unshadowed tracks, shadowed tracks, and starting tracks).

While much focus in the literature has been given to the discovery potential of ultra-high-energy neutrino detectors, detailed flavor composition measurements have remained largely unexplored. Indeed, distinguishing between the signals of different flavors with novel radio, optical, and Earth-skimming technologies requires specialized techniques, so many analyses avoid full flavor separation. In this work, we showcase how a complete flavor measurement at 95\% C.L. can be achieved at energies $10^2$ - $10^4\,\mathrm{PeV}$ by combining the three technologies.

The first strategy uses measurement forecasts of in-ice radio neutrino telescopes by Coleman et al.~\cite{Coleman:2024scd}.
Electron neutrino charged-current interactions are isolated using a neural network trained to recognize the temporally elongated radio emission characteristic of the Landau--Pomeranchuk--Migdal (LPM) effect. 
Further, the muon and tau-neutrino fractions ($\nu_\mu$, $\nu_\tau$) can be jointly constrained through the identification of multiple shower signatures produced by secondary muons and taus.
While this enables sensitivity to all three flavors, it leaves an intrinsic degeneracy between $\nu_\mu$ and $\nu_\tau$.

The second strategy resolves this limitation by integrating measurements from Earth-skimming detectors, which, excluding the Glashow resonance, are uniquely sensitive to $\nu_\tau$. This complementary channel provides a direct determination of the $\nu_\tau$ contribution to the diffuse UHE flux, thereby breaking the $\nu_\mu$--$\nu_\tau$ degeneracy that is inherent to a radio-based analysis.

Combined with in-ice optical Cherenkov, these three strategies enable a full reconstruction of the neutrino flavor composition above $100\,\mathrm{PeV}$. Although $\nu_\tau$ sensitivity has previously been demonstrated in isolation~\cite{Testagrossa:2023ukh}, here we present the first forecast of a simultaneous measurement of $\nu_e$, $\nu_\mu$, and $\nu_\tau$  within a unified framework. This projected measurement, showcased in Figure~\ref{fig:uhe_flavor}, carries information about both the production mechanisms at the source and the effects of neutrino propagation~\cite{Arguelles:2015dca}.

The rest of this article is structured as follows. Section~\ref{sec:source} introduces competing hypotheses for neutrino flavor composition at production. Section~\ref{sec:stats} explains the statistical analysis and generation of mock event samples. Sections~\ref{sec:skimming} and~\ref{sec:radio} detail the flavor reconstruction techniques for Earth-skimming arrays and in-ice radio detectors, respectively. Finally, Section~\ref{sec:transition} assesses the potential of this ultra-high-energy flavor framework to uncover energy-dependent variations in the neutrino flavor composition at production, under the assumption of perfect knowledge of neutrino oscillation parameters.

\begin{widetext}
\begin{center}
    \includegraphics[width=\linewidth]{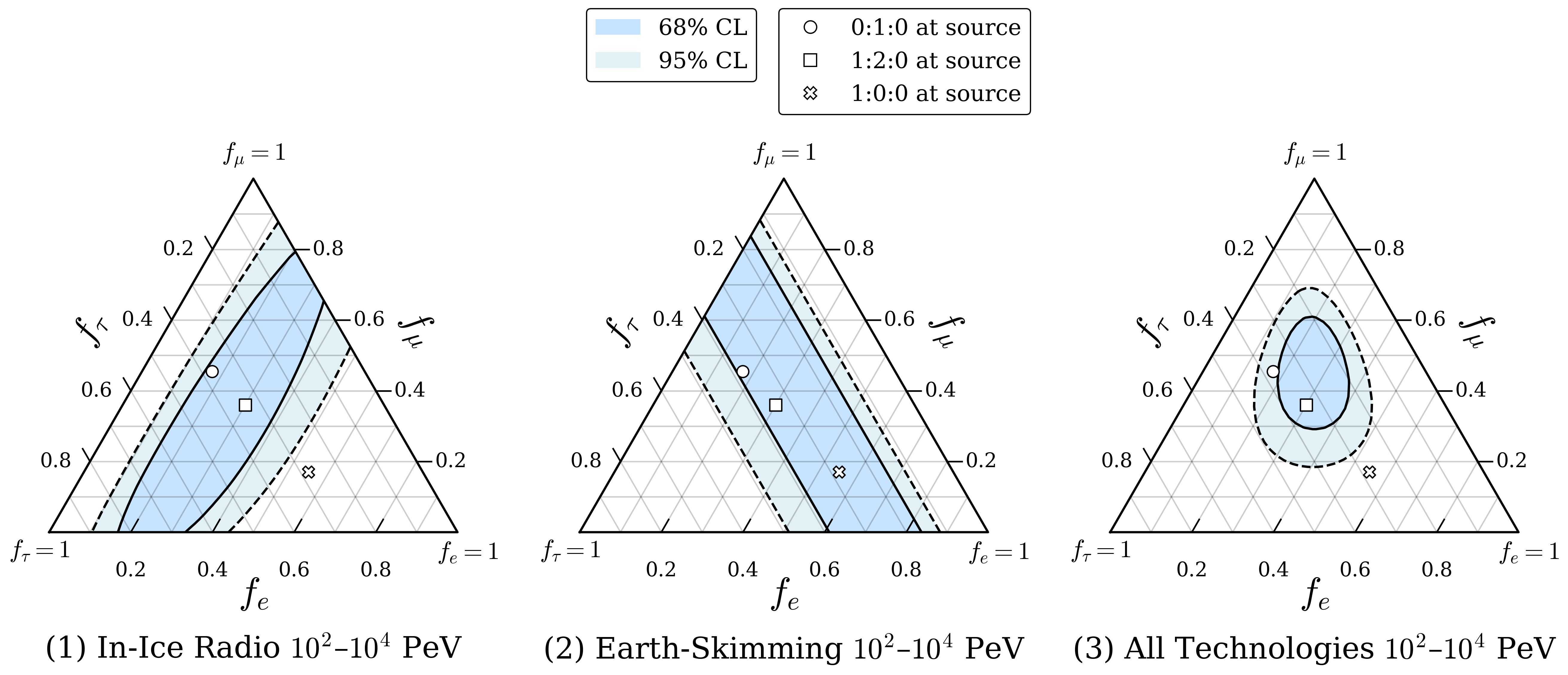}
        \captionof{figure}{Flavor sensitivity of ultra-high-energy neutrinos in the range $10^{2}$--$10^{4}\,\mathrm{PeV}$ over 10 years of exposure.
    The injected source ratio $(1:2:0)_S$ from pion decay evolves to $(0.30:0.36:0.34)$ at Earth. The flux model combines the IceCube 9.5-year muon-neutrino measurement with a ``realistic'' cosmogenic UHECR flux, averaging over fits to UHECR data from Telescope Array (TA) and AUGER Cosmogenic assuming 10\% proton content of cosmic rays. \cite{Coleman:2024scd}
    Dashed lines denote 95\% C.L. and solid lines 68\% C.L., obtained from a likelihood-ratio analysis. Systematic uncertainties are not modelled, since statistical uncertainties dominate. Panels show: (\textbf{left}) in-ice radio detectors (ARIANNA, ARA, RNO-G, IceCube-Gen2); (\textbf{middle}), Earth-skimming experiments (TAMBO-5k, POEMMA, Trinity, AUGER, GRAND200k); and (\textbf{right}) the combined all-experiment flavor sensitivity. 
    We forecast a UHE flavor measurement of $\nu_e$, $\nu_\mu$, and $\nu_\tau$. }
    \label{fig:uhe_flavor}
\end{center}
\clearpage
\end{widetext}

\section{Neutrino Flavor Composition at the Source}
\label{sec:source}

Three source flavor compositions are considered in this work. When ultra-high-energy cosmic rays interact with matter around their acceleration sources, charged pions are produced and decay via  
\(\pi^+ \rightarrow \mu^+ + \nu_\mu\), followed by \(\mu^+ \rightarrow e^+ + \nu_e + \bar{\nu}_\mu\), giving the standard pion-decay flavor composition \((f_{e,S},f_{\mu,S},f_{\tau,S})=(1/3,\,2/3,\,0)_S\). In sources with strong magnetic fields, secondary muons cool through synchrotron radiation before decaying, suppressing the electron-neutrino component above a critical energy of \(\sim2\,\mathrm{PeV}\), resulting in the muon-damped composition \((0,1,0)_S\). Alternatively, beta decay of neutrons and neutron-rich isotopes produces a pure \(\bar{\nu}_e\) flux with composition \((1,0,0)_S\) \cite{Coleman:2024scd}. Figure~\ref{fig:uhe_flavor} shows that this neutron-decay hypothesis is excluded at 95\% C.L. by in-ice radio experiments.

\section{Statistical Analysis}
\label{sec:stats}

Event samples for in-ice radio, Cherenkov, and Earth-skimming detectors are generated under benchmark assumptions for the all-flavor flux, $\Phi_{\nu,\text{all}}$, which includes both a cosmogenic and astrophysical component (see Appendix~\ref{app:flux} for further details). These samples constitute the observed data, with observable $N_{\nu,\text{all},i}$ per energy bin $i$. 

Detector response is encoded in the flavor-dependent effective area $A_{\alpha}(E)$. For each Earth-skimming detector, $A_{\alpha}(E)$ is obtained from projected sensitivities to the all-flavor astrophysical neutrino flux. For in-ice radio detectors, it is instead derived from the per-flavor effective volumes $V_{\alpha}(E)$.

Given the all-flavor flux $\Phi_{\nu,\text{all}}$ and the flavor composition $f_{\alpha,\oplus}$ at Earth, the differential event rate for flavor $\alpha$ after an exposure time $T$ is
\begin{align}
\frac{dN_\alpha}{dE} 
= 4\pi T \, f_{\alpha,\oplus}(E)\,\Phi_{\nu,\text{all}}(E)\, A_{\alpha}(E),
\label{eq:event_spectrum}
\end{align}
where the factor $4\pi$ accounts for the full-sky integration. Forecasts shown in this article will assume an exposure of $T=10$ years unless stated otherwise.

To infer the most likely flavor composition $(f_{e,\oplus}, f_{\mu,\oplus}, f_{\tau,\oplus})$, the observed event sample is compared to test samples generated under different hypotheses. Since the flavor fractions satisfy $f_{\tau,\oplus} = 1 - f_{e,\oplus} - f_{\mu,\oplus}$, only $f_{e,\oplus}$ and $f_{\mu,\oplus}$ are treated as free parameters.  

Statistical fluctuations in detected event counts are modeled as Poisson distributions. The log-likelihood function, constructed from Poisson probabilities for each observable, reduces to
\begin{align}
\ln \mathcal{L}(f_\oplus)
= \sum_i \big[ N_i^{\text{obs}} \ln \mu_i(f_\oplus) - \mu_i(f_\oplus) \big],
\end{align}
where $\mu_i(f_\oplus)$ denotes the expected number of events in bin $i$ for the competing flavor hypothesis $f_\oplus$, and $N_i^{\text{obs}}$ is the number of events observed in data. Modifications to the likelihood are introduced in Sections~\ref{sec:skimming} and~\ref{sec:radio} to account for specialized detection channels in Earth-skimming and radio experiments.

\section{Earth-Skimming Flavor Sensitivity}
\label{sec:skimming}

For Earth-skimming $\nu_\tau$ detectors, average effective areas are reconstructed from the sensitivity estimates $S(E_\nu)$ \cite{Ackermann:2022rqc}. The average effective area is 
\begin{equation}
A_{\mathrm{avg}}(E_\nu) = \frac{2.44\,E_\nu}{\Omega\,S(E_\nu)\,T\,\ln 10},
\end{equation}
where $\Omega$ is the solid-angle coverage and the factor of $2.44$ events arises from a background-free prescription~\cite{Feldman:1997qc}.

TAMBO and Trinity experiments are also sensitive to a $\bar{\nu}_e$ flux. Throughout most of the energy range, the signal comes from $\nu_\tau$-initiated showers, but around $6.3\,\mathrm{PeV}$ there is an important $\bar{\nu}_e$ contribution due to the Glashow resonance. Electron neutrinos $\bar{\nu}_e$ interact with electrons to produce $W^-$ bosons that subsequently decay into tau leptons. The average ${\nu}_\tau$ and $\bar{\nu}_e$ effective areas for TAMBO and Trinity are calculated from Neutrino Acceptance estimates $A_{acc}(E_\nu)$ \cite{Ding:2026wjp, Huang:2019hgs}. Trinity's air-shower imaging geometry (with a horizontal field of view of 60° and a vertical field of view of 5°) differs substantially to TAMBO's ground-array configuration (spanning 120° by 30°).

\begin{figure}[H]
    \centering
    \includegraphics[width=0.8\linewidth]{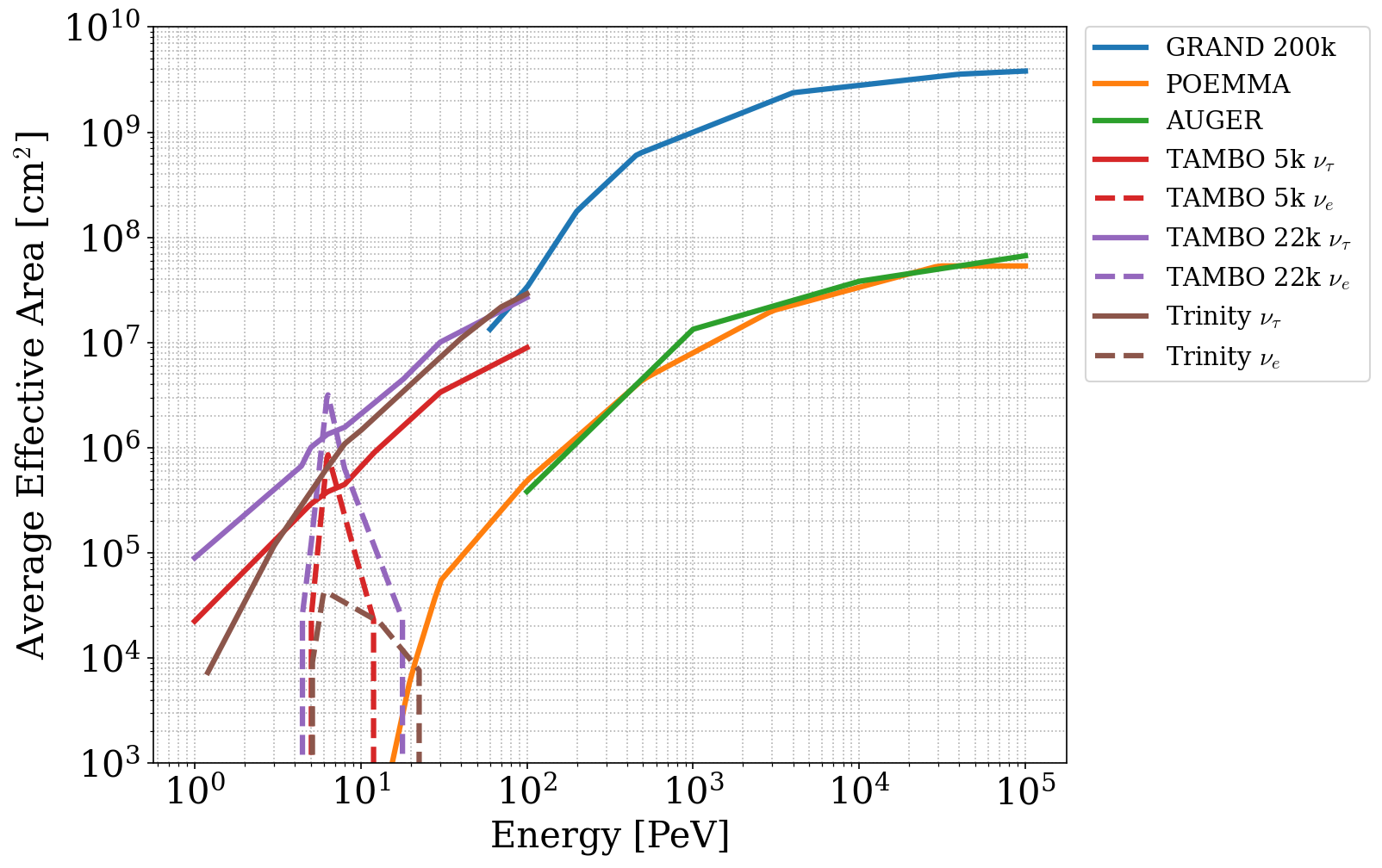}
    \caption{
Average earth-skimming effective areas (cm$^{2}$) for TAMBO-5k, TAMBO-22k, POEMMA, Trinity, AUGER, and GRAND200k ~\cite{Stepanoff:2025vys, Ackermann:2022rqc}.
Solid lines show the contribution from the primary $\nu_\tau$ channel, while dashed lines indicate the $\bar{\nu}_e$ contribution from the Glashow resonance.
}
    \label{fig:effareas}
\end{figure}

A comparison of the average effective areas of TAMBO-5k, TAMBO-22k, POEMMA, Trinity, AUGER, and GRAND200k is shown in Figure~\ref{fig:effareas}.

The occurrence of Glashow-resonance interactions is modeled as a binomial process within each analysis bin $i$. Out of a total of $N_{\mathrm{all},i}$ observed events in the bin, a subset $N^{\mathrm{obs}}_{G,i}$ are identified as Glashow-tagged. The probability of this outcome, given model parameters $\theta$, is
\begin{equation}
P_{G,i}(\theta) =
\mathcal{B}\!\left(N^{\mathrm{obs}}_{G,i};\, N_{\mathrm{all},i},\, p_{G,i}(\theta)\right),
\end{equation}
where
\begin{equation}
\mathcal{B}(k;n,p) \equiv \binom{n}{k} p^{k} (1-p)^{\,n-k},
\end{equation}
with $n=N_{\mathrm{all},i}$ the number of trials, $k=N^{\mathrm{obs}}_{G,i}$ the number of observed Glashow events, and $p_{G,i}(\theta)$ the normalized per-event probability of producing a $\bar{\nu}_e$ channel signature. See Appendix~\ref{app:skimming}.

The overall likelihood across all bins is then the product of independent binomials, equivalently the sum of log-likelihood terms,
\begin{equation}
\log \mathcal{L}_G(\theta) =
\sum_i \log \mathcal{B}\!\left(
N^{\mathrm{obs}}_{G,i};\,
N_{\mathrm{all},i},\,
p_{G,i}(\theta)
\right).
\end{equation}

\section{In-Ice Radio Flavor Sensitivity}
\label{sec:radio}

In-ice radio detector response is estimated from IceCube-Gen2 simulated effective volumes computed with \texttt{NuRadioMC}\cite{Coleman:2024scd}. We assume that the detector response of other radio experiments is similar to that of IceCube-Gen2, and effective volumes are scaled accordingly. Volumes are direction-averaged and include Earth attenuation. At 100PeV, for example, the $\nu_e$ CC effective volume dominates, exceeding others by roughly a factor of 5.

From effective volumes $V_{\alpha}^{\rm CC}$ and $V_{\alpha}^{\rm NC}$, one derives effective areas,
\begin{align}
A_{\alpha}^{\rm CC} &= \frac{V_{\alpha}^{\rm CC}}{\lambda_{\alpha}^{\rm CC}}, \quad 
A_{\alpha}^{\rm NC} = \frac{V_{\alpha}^{\rm NC}}{\lambda_{\alpha}^{\rm NC}},
\end{align}

where $\lambda_{\alpha}^{\rm CC} = (\sigma_{\nu N}^{\rm CC} n_N)^{-1}$, $\lambda_{\alpha}^{\rm NC} = (\sigma_{\nu N}^{\rm NC} n_N)^{-1}$ are the CC and NC interaction lengths in ice and $n_N$ the nucleon number density in ice. Here $\sigma_{\nu N}$ are the UHE neutrino-nucleon cross sections \cite{Formaggio:2012cpf}.

We consider three observables in this analysis:  

1. \textbf{Total event spectrum $N_{\nu,\text{all},i}(\theta)$}, as in the previous section.

2. \textbf{Electron-neutrino $\nu_e$ CC spectrum $N_{\text{ML},i}(\theta)$}, events classified as charged-current $\nu_e$ interactions.

The LPM effect discriminates charged-current (CC) interactions of $\nu_e$ from other neutrino flavors, as demonstrated in prior work~\cite{Barwick:2022vqt}.

3. \textbf{Multi-shower spectrum $N_{\text{mult},i}(\theta)$}, events identified as multi-shower signatures from charged-current interactions of $\nu_\mu$ and $\nu_\tau$. A hadronic shower is produced at the $\nu N$ interaction vertex, and the outgoing muon or tau lepton can generate additional high-energy secondary showers while propagating through the ice. These arise from stochastic energy losses of the muon and, in the case of the tau, also from its decay~\cite{Garcia-Fernandez:2020dhb}.  

Combining the $\nu_e$ CC classification channel and the multi-shower channel, the full likelihood expression used in this work is  

\begin{align}
\mathcal{L}_{\text{ML+mult}}(\theta) = \prod_{i=1}^{N_E} \Big[
&\mathcal{P}\!\left(N_{\nu,\text{all},i}^{\text{obs}};\, N_{\nu,\text{all},i}(\theta)\right) \nonumber \\
&\times P_{\text{ML},i}(\theta)\, P_{\text{mult},i}(\theta) \Big].
\end{align}

Here $N_E$ is the number of energy bins. This likelihood calculation provides joint sensitivity to $\nu_e$ via LPM-based classification and to $\nu_\mu+\nu_\tau$ via multi-shower signatures. Figure~\ref{fig:figureXXX} displays the flavor sensitivity obtained from each channel individually, as well as the combined result for an exposure period of 10 years. We consider three cosmogenic flux hypotheses: low, mid, and high, based on the IceCube 9.5-year muon-neutrino measurement combined with a cosmogenic component. The panels show (\textbf{left}) the low-flux cosmogenic forecast based on AUGER data, (\textbf{middle}) the geometric mean of the low- and high-flux cosmogenic contributions; and (\textbf{right}) the high-flux cosmogenic forecast based on Telescope Array data~\cite{Coleman:2024scd}.

The energy-dependent true and false-positive rates for $\nu_e$ charged-current (CC) events, together with the fraction of ultra-high-energy
$\nu_\mu$- and $\nu_\tau$-initiated CC interactions identified as multi-shower events, are shown in Appendix~\ref{app:radio} (Fig.~\ref{fig:fp_rates}). Appendix~\ref{app:radio} also provides additional details of the likelihood calculation.

\begin{widetext}
    \centering
    \includegraphics[width=\linewidth]{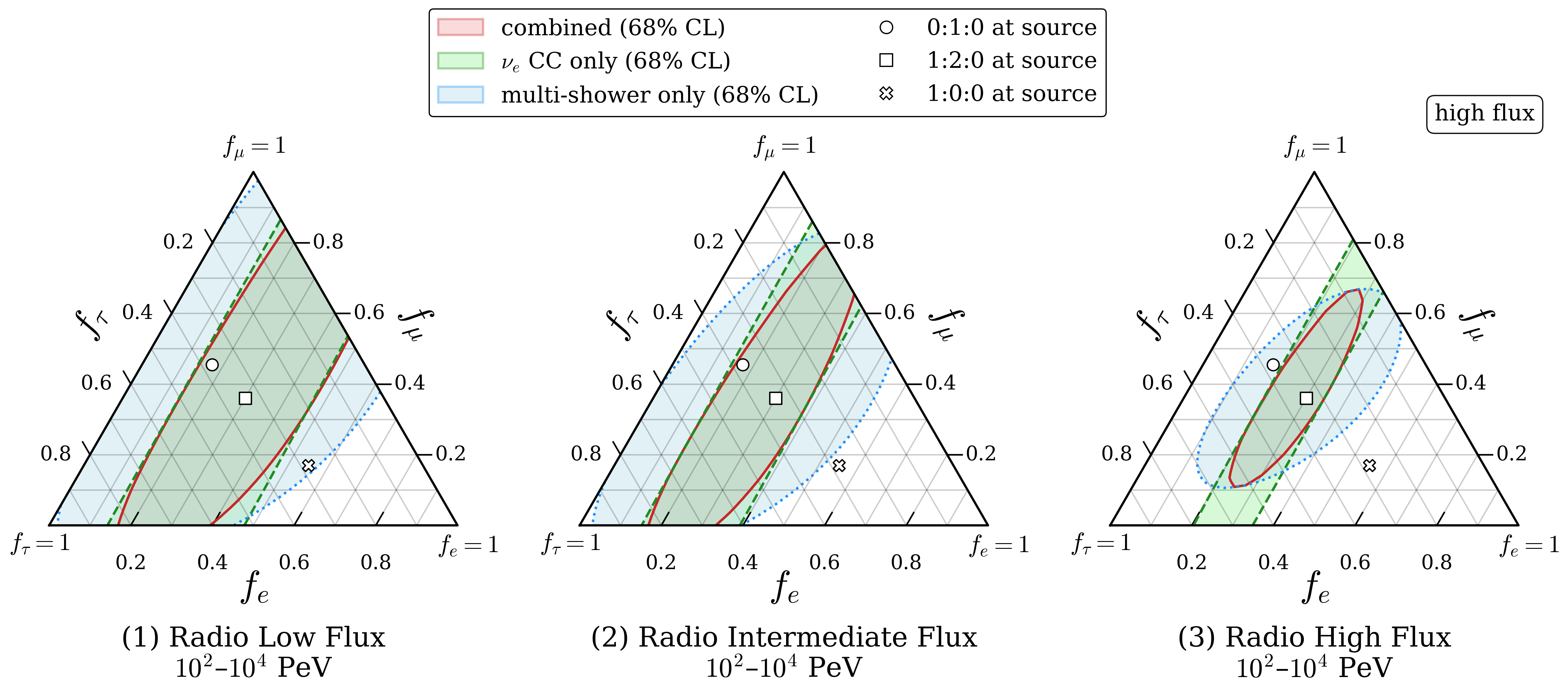}
    \captionof{figure}{Per-channel flavor sensitivity of in-ice radio experiments for $10^{2}$--$10^{4}\,\mathrm{PeV}$ neutrinos under low, mid, and high-flux scenarios. Contours are drawn at 68\%~confidence level over an exposure period of 10 years. Systematic uncertainties are not modelled, since statistical uncertainties dominate. The plot shows contributions from the $\nu_e$ charged-current (CC) channel, the multi-shower channel, and the combined sensitivity.  The injected source flavor ratio $(1\!:\!2\!:\!0)_S$ from pion decay evolves to $(0.30\!:\!0.36\!:\!0.34)$ at Earth. Systematic uncertainties are not included.
}
    \label{fig:figureXXX}
\end{widetext}

\section{Measuring an Energy-Dependent Flavor Transition at Production}
\label{sec:transition}

As an application of the three-technology framework, we study sensitivity to an energy-dependent neutrino flavor transition at production. We assume a high-flux scenario with 15 years of exposure and use IceCube-Gen2 optical Cherenkov effective areas from the \texttt{TOISE} code~\cite{vanSanten:2022wss}. The neutrino flavor fractions at Earth, \(f_{e,\oplus}\) and \(f_{\mu,\oplus}\), are extrapolated back to the source assuming standard neutrino PMNS oscillations. Averaged oscillations give transition probabilities of \( P_{\alpha\beta}=\sum_i |U_{\alpha i}|^2 |U_{\beta i}|^2\)
such that the flavor composition at Earth is
\( f_{\alpha,\oplus}=\sum_{\beta=e,\mu,\tau} P_{\beta\alpha} f_{\beta,S}.\)

\begin{widetext}
    \centering
    \includegraphics[width=\textwidth]{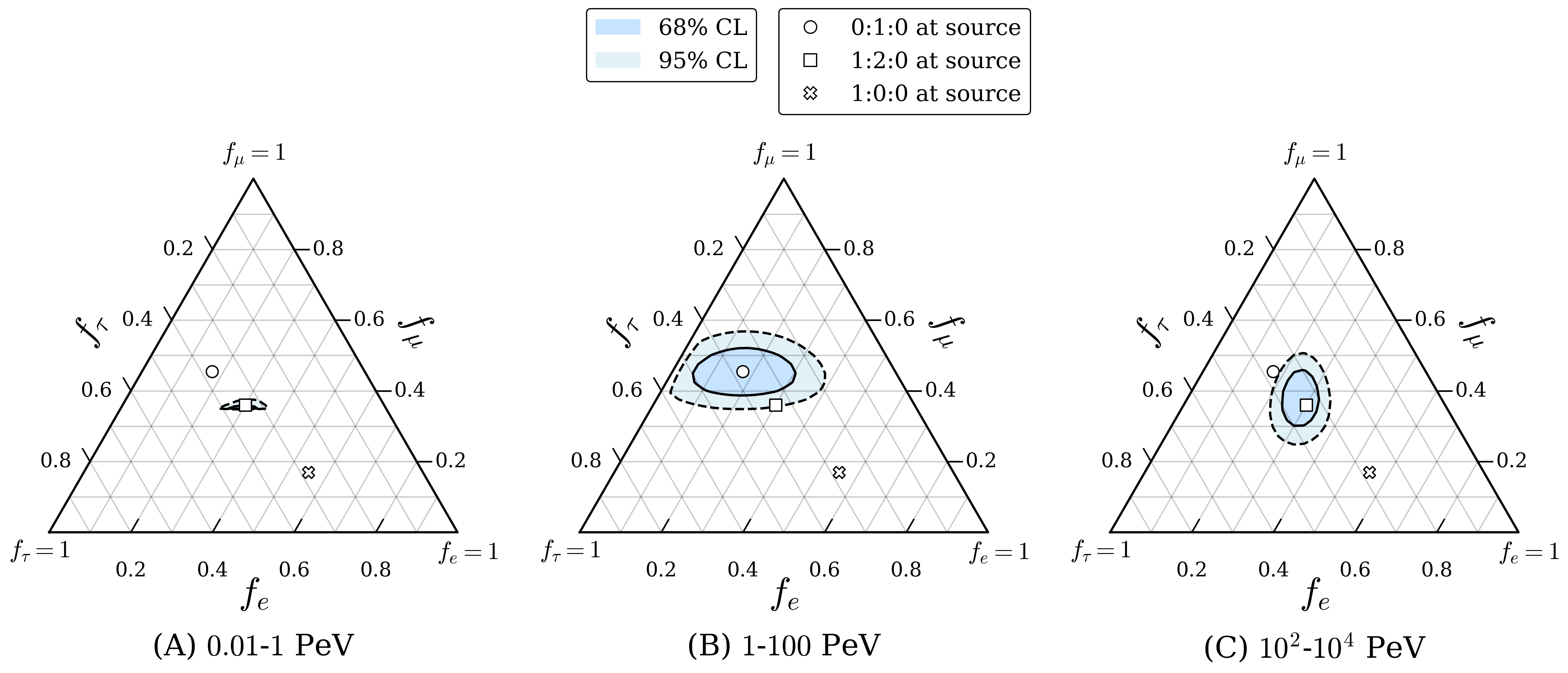}\\[3pt]
    \includegraphics[width=\textwidth]{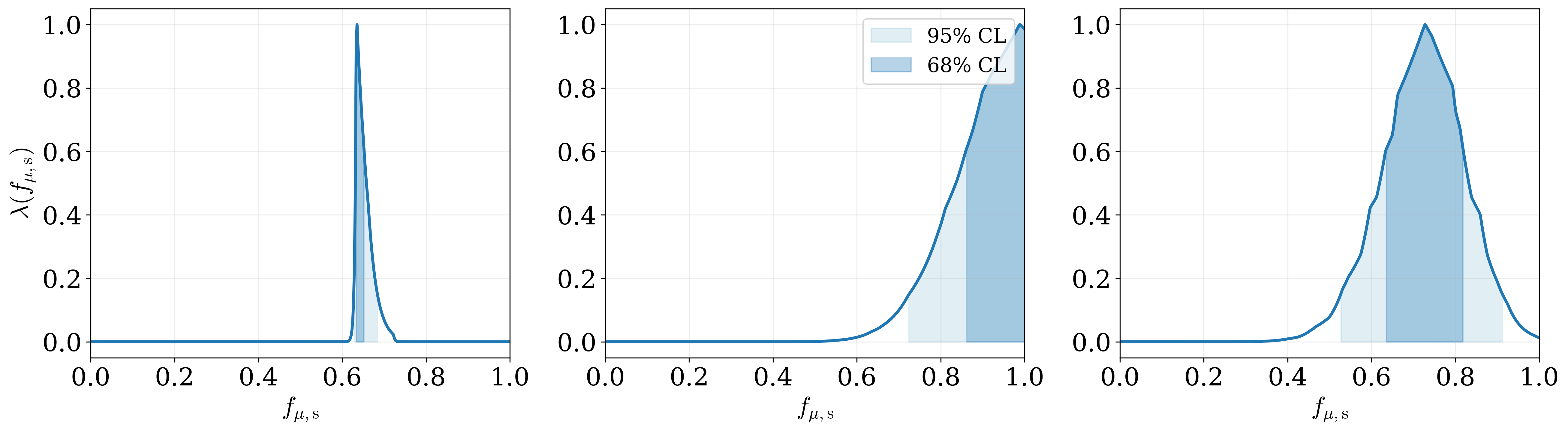}
    \captionof{figure}{
Energy-dependent flavor composition forecasts spanning TeV--EeV energies.
The upper panel shows flavor triangles from a 15-year, all-experiment integration,
binned in the ranges $0.01$--$1$, $1$--$100$, and $100$--$1000\,\mathrm{PeV}$.
The injected source distribution transitions from $(1\!:\!2\!:\!0)_S$ to $(0\!:\!1\!:\!0)_S$ and back to $(1\!:\!2\!:\!0)_S$ across the three energy bins. The lower panels show the  distributions for the $\nu_\mu$ source fraction, assuming standard oscillations. Dark and light shaded regions indicate the 68\% and 95\% confidence levels, respectively. Whilst statistical uncertainties dominate in the middle and right panels, systematic detector effects become important in the left panel. We leave a full treatment of these for a future study.
}
    \label{fig:muonfrac}
\end{widetext}

We adopt the NuFit~5.0 best-fit oscillation parameters for normal mass ordering~\cite{Esteban:2020cvm} (see also the updated NuFit~6.0 fit~\cite{Esteban:2024eli}) and assume perfect knowledge of mixing parameters, consistent with projected 2040 precision. Under these assumptions, the framework predicts sensitivity to a transition from pion-decay \((1/3,2/3,0)_S\) to muon-damped \((0,1,0)_S\) production at the 95\% C.L. around PeV energies, with discrimination at 68\% C.L. at \(100\,\mathrm{PeV}\) (Fig.~\ref{fig:muonfrac}). One major limitation of the method is the uncertainty in the UHE and cosmogenic neutrino flux at present. 

This line of inquiry is important for two reasons. First, it provides a probe of magnetic field intensity inside UHE neutrino sources. Muon damping occurs when synchrotron cooling suppresses muon decay before neutrino production, and becomes important above energies of 
\[
E \gtrsim 2\times10^{9}\,\left(\Gamma/B\right)\,\mathrm{GeV},
\]
where \(\Gamma\) is the bulk Lorentz factor and \(B\) is the magnetic-field strength. An energy-dependent transition at UHE provides sensitivity to source magnetic fields down to the \(\mathcal{O}(1\,\mathrm{G})\) scale~\cite{Coleman:2024scd}. Second, flavor measurements may help distinguish cosmogenic from UHE neutrinos. Since extragalactic magnetic fields are expected to be weak, cosmogenic neutrinos should retain the standard pion-decay flavor composition at production, transitioning from \((0,1,0)_S\) back to \((1/3,2/3,0)_S\).

\section{Conclusions and Outlook}
\label{sec:conclusions}

Because their Standard Model interaction cross section is small, neutrinos can probe otherwise inaccessible regions such as the interiors of stars and astrophysical accelerators. 
Here, we highlight the discovery potential of a coordinated multi-technology program for studying ultra-high-energy neutrino flavor physics.
We forecast a full flavor measurement over 10 years of exposure, under the assumption of various experiments been deployed, and use it to investigate energy-dependent transitions in the source flavor composition. Such measurements are interesting for probing Beyond-the-Standard-Model scenarios such as sterile-neutrino effects at production, dark matter interactions during propagation, or non-standard interactions in Earth matter at detection~\cite{Arguelles:2022tki}. Given these promising results and the limited prior art, we think neutrino flavor measurements should form an integral component of the design and strategy of upcoming Earth-skimming, radio-based, and optical-Cherenkov neutrino observatories.

\section*{Acknowledgements}
C.A.A. are supported by the Faculty of Arts and Sciences of Harvard University, the National Science Foundation (NSF), the John Templeton Foundation, the Research Corporation for Science Advancement, and the David \& Lucile Packard Foundation.
W.G.T. is supported by the John Templeton Foundation (Grant \#63651).
This publication was made possible through the support of Grant 63651 from the John Templeton Foundation. The opinions expressed in this publication are those of the authors and do not necessarily reflect the views of the John Templeton Foundation.

\bibliographystyle{apsrev4-2}
\bibliography{refs}

@article{Berezinsky:1969erk,
    author = "Berezinsky, V. S. and Zatsepin, G. T.",
    title = "{Cosmic rays at ultrahigh-energies (neutrino?)}",
    doi = "10.1016/0370-2693(69)90341-4",
    journal = "Phys. Lett. B",
    volume = "28",
    pages = "423--424",
    year = "1969"
}

@article{KM3NeT:2025npi,
    author = "Aiello, S. and others",
    collaboration = "KM3NeT",
    title = "{Observation of an ultra-high-energy cosmic neutrino with KM3NeT}",
    doi = "10.1038/s41586-024-08543-1",
    journal = "Nature",
    volume = "638",
    number = "8050",
    pages = "376--382",
    year = "2025",
    note = "[Erratum: Nature 640, E3 (2025)]"
}

@article{ANITA:2020gmv,
    author = "Gorham, P. W. and others",
    collaboration = "ANITA",
    title = "{Unusual Near-Horizon Cosmic-Ray-like Events Observed by ANITA-IV}",
    eprint = "2008.05690",
    archivePrefix = "arXiv",
    primaryClass = "astro-ph.HE",
    doi = "10.1103/PhysRevLett.126.071103",
    journal = "Phys. Rev. Lett.",
    volume = "126",
    number = "7",
    pages = "071103",
    year = "2021"
}

@article{RNO-G:2020rmc,
    author = "Aguilar, J. A. and others",
    collaboration = "RNO-G",
    title = "{Design and Sensitivity of the Radio Neutrino Observatory in Greenland (RNO-G)}",
    eprint = "2010.12279",
    archivePrefix = "arXiv",
    primaryClass = "astro-ph.IM",
    doi = "10.1088/1748-0221/16/03/P03025",
    journal = "JINST",
    volume = "16",
    number = "03",
    pages = "P03025",
    year = "2021",
    note = "[Erratum: JINST 18, E03001 (2023)]"
}

@online{POEMMA:2020ykm,
    author = "Olinto, A. V. and others",
    collaboration = "POEMMA",
    title = "{The POEMMA (Probe of Extreme Multi-Messenger Astrophysics) observatory}",
    eprint = "2012.07945",
    archivePrefix = "arXiv",
    primaryClass = "astro-ph.IM",
    doi = "10.1088/1475-7516/2021/06/007",
    journal = "JCAP",
    volume = "06",
    pages = "007",
    year = "2021"
}

@article{GRAND:2018iaj,
    author = "{\'A}lvarez-Mu{\~n}iz, Jaime and others",
    collaboration = "GRAND",
    title = "{The Giant Radio Array for Neutrino Detection (GRAND): Science and Design}",
    eprint = "1810.09994",
    archivePrefix = "arXiv",
    primaryClass = "astro-ph.HE",
    doi = "10.1007/s11433-018-9385-7",
    journal = "Sci. China Phys. Mech. Astron.",
    volume = "63",
    number = "1",
    pages = "219501",
    year = "2020"
}

@article{Testagrossa:2023ukh,
    author = "Testagrossa, Federico and Fiorillo, Damiano F. G. and Bustamante, Mauricio",
    title = "{Two-detector flavor sensitivity to ultrahigh-energy cosmic neutrinos}",
    eprint = "2310.12215",
    archivePrefix = "arXiv",
    primaryClass = "astro-ph.HE",
    doi = "10.1103/PhysRevD.110.083026",
    journal = "Phys. Rev. D",
    volume = "110",
    number = "8",
    pages = "083026",
    year = "2024"
}

@article{IceCube-Gen2:2021rkf,
    author = "Abbasi, Rasha and others",
    collaboration = "IceCube-Gen2",
    title = "{Sensitivity studies for the IceCube-Gen2 radio array}",
    eprint = "2107.08910",
    archivePrefix = "arXiv",
    primaryClass = "astro-ph.HE",
    reportNumber = "PoS-ICRC2021-1183",
    doi = "10.22323/1.395.1183",
    journal = "PoS",
    volume = "ICRC2021",
    pages = "1183",
    year = "2021"
}

@article{Feldman:1997qc,
    author = "Feldman, Gary J. and Cousins, Robert D.",
    title = "{A Unified approach to the classical statistical analysis of small signals}",
    eprint = "physics/9711021",
    archivePrefix = "arXiv",
    reportNumber = "HUTP-97-A096",
    doi = "10.1103/PhysRevD.57.3873",
    journal = "Phys. Rev. D",
    volume = "57",
    pages = "3873--3889",
    year = "1998"
}

@article{Ackermann:2022rqc,
    author = "Ackermann, Markus and others",
    title = "{High-energy and ultra-high-energy neutrinos: A Snowmass white paper}",
    eprint = "2203.08096",
    archivePrefix = "arXiv",
    primaryClass = "hep-ph",
    doi = "10.1016/j.jheap.2022.08.001",
    journal = "JHEAp",
    volume = "36",
    pages = "55--110",
    year = "2022"
}

@article{Formaggio:2012cpf,
    author = "Formaggio, J. A. and Zeller, G. P.",
    title = "{From eV to EeV: Neutrino Cross Sections Across Energy Scales}",
    eprint = "1305.7513",
    archivePrefix = "arXiv",
    primaryClass = "hep-ex",
    reportNumber = "FERMILAB-PUB-12-785-E",
    doi = "10.1103/RevModPhys.84.1307",
    journal = "Rev. Mod. Phys.",
    volume = "84",
    pages = "1307--1341",
    year = "2012"
}

@article{Garcia-Fernandez:2020dhb,
    author = "Garc{\'\i}a-Fern{\'a}ndez, Daniel and Nelles, Anna and Glaser, Christian",
    title = "{Signatures of secondary leptons in radio-neutrino detectors in ice}",
    eprint = "2003.13442",
    archivePrefix = "arXiv",
    primaryClass = "astro-ph.HE",
    reportNumber = "DESY-20-055",
    doi = "10.1103/PhysRevD.102.083011",
    journal = "Phys. Rev. D",
    volume = "102",
    number = "8",
    pages = "083011",
    year = "2020"
}

@misc{Barwick:2022vqt,
    author = "Barwick, Steven W. and Glaser, Christian",
    editor = "Fazi, Giovanni G.",
    title = "{Chapter 6: Radio Detection of High Energy Neutrinos in Ice}",
    eprint = "2208.04971",
    archivePrefix = "arXiv",
    primaryClass = "astro-ph.IM",
    doi = "10.1142/9789811282645_0006",
    pages = "237--302",
    year = "2023", 
}

@article{vanSanten:2022wss,
    author = "van Santen, Jakob and Clark, Brian A. and Halliday, Rob and Hallmann, Steffen and Nelles, Anna",
    title = "{toise: a framework to describe the performance of high-energy neutrino detectors}",
    eprint = "2202.11120",
    archivePrefix = "arXiv",
    primaryClass = "astro-ph.IM",
    doi = "10.1088/1748-0221/17/08/T08009",
    journal = "JINST",
    volume = "17",
    number = "08",
    pages = "T08009",
    year = "2022"
}

@article{Esteban:2020cvm,
    author = "Esteban, Ivan and Gonzalez-Garcia, M. C. and Maltoni, Michele and Schwetz, Thomas and Zhou, Albert",
    title = "{The fate of hints: updated global analysis of three-flavor neutrino oscillations}",
    eprint = "2007.14792",
    archivePrefix = "arXiv",
    primaryClass = "hep-ph",
    reportNumber = "IFT-UAM/CSIC-112, YITP-SB-2020-21",
    doi = "10.1007/JHEP09(2020)178",
    journal = "JHEP",
    volume = "09",
    pages = "178",
    year = "2020"
}

@article{Arguelles:2015dca,
    author = {Arg{\"u}elles, Carlos A. and Katori, Teppei and Salvado, Jordi},
    title = "{New Physics in Astrophysical Neutrino Flavor}",
    eprint = "1506.02043",
    archivePrefix = "arXiv",
    primaryClass = "hep-ph",
    doi = "10.1103/PhysRevLett.115.161303",
    journal = "Phys. Rev. Lett.",
    volume = "115",
    pages = "161303",
    year = "2015"
}

@article{IceCube:2013low,
    author = "Aartsen, M. G. and others",
    collaboration = "IceCube",
    title = "{Evidence for High-Energy Extraterrestrial Neutrinos at the IceCube Detector}",
    eprint = "1311.5238",
    archivePrefix = "arXiv",
    primaryClass = "astro-ph.HE",
    doi = "10.1126/science.1242856",
    journal = "Science",
    volume = "342",
    pages = "1242856",
    year = "2013"
}

@article{Allison:2011wk,
    author = "Allison, P. and others",
    title = "{Design and Initial Performance of the Askaryan Radio Array Prototype EeV Neutrino Detector at the South Pole}",
    eprint = "1105.2854",
    archivePrefix = "arXiv",
    primaryClass = "astro-ph.IM",
    doi = "10.1016/j.astropartphys.2011.11.010",
    journal = "Astropart. Phys.",
    volume = "35",
    pages = "457--477",
    year = "2012"
}

@article{Persichilli:2019wyh,
    author = "Persichilli, Christopher Robert",
    collaboration = "ARIANNA",
    title = "{Performance of the ARIANNA pilot array, and implications for the next generation of UHE neutrino detectors}",
    doi = "10.22323/1.358.0980",
    journal = "PoS",
    volume = "ICRC2019",
    pages = "980",
    year = "2020"
}

@article{Coleman:2024scd,
    author = "Coleman, Alan and Ericsson, Oscar and Glaser, Christian and Bustamante, Mauricio",
    title = "{Flavor composition of ultrahigh-energy cosmic neutrinos: Measurement forecasts for in-ice radio-based EeV neutrino telescopes}",
    eprint = "2402.02432",
    archivePrefix = "arXiv",
    primaryClass = "astro-ph.HE",
    doi = "10.1103/PhysRevD.110.023044",
    journal = "Phys. Rev. D",
    volume = "110",
    number = "2",
    pages = "023044",
    year = "2024"
}

@article{Stepanoff:2025vys,
    author = "Stepanoff, Sofia",
    collaboration = "Trinity",
    title = "{Status of the Trinity PeV Neutrino Observatory}",
    eprint = "2509.18236",
    archivePrefix = "arXiv",
    primaryClass = "astro-ph.IM",
    doi = "10.22323/1.501.1188",
    journal = "PoS",
    volume = "ICRC2025",
    pages = "1188",
    year = "2025"
}

@online{TAMBO:2025jio,
    author = {Arg{\"u}elles, Carlos A. and others},
    collaboration = "TAMBO",
    title = "{TAMBO: A Deep-Valley Neutrino Observatory}",
    eprint = "2507.08070",
    archivePrefix = "arXiv",
    primaryClass = "astro-ph.HE",
    month = "7",
    year = "2025", 
}

@article{IceCube:2018fhm,
    author = "Aartsen, M. G. and others",
    collaboration = "IceCube",
    title = "{Differential limit on the extremely-high-energy cosmic neutrino flux in the presence of astrophysical background from nine years of IceCube data}",
    eprint = "1807.01820",
    archivePrefix = "arXiv",
    primaryClass = "astro-ph.HE",
    doi = "10.1103/PhysRevD.98.062003",
    journal = "Phys. Rev. D",
    volume = "98",
    number = "6",
    pages = "062003",
    year = "2018"
}

@article{Arguelles:2022tki,
    author = {Arg{\"u}elles, C. A. and others},
    title = "{Snowmass white paper: beyond the standard model effects on neutrino flavor: Submitted to the proceedings of the US community study on the future of particle physics (Snowmass 2021)}",
    eprint = "2203.10811",
    archivePrefix = "arXiv",
    primaryClass = "hep-ph",
    doi = "10.1140/epjc/s10052-022-11049-7",
    journal = "Eur. Phys. J. C",
    volume = "83",
    number = "1",
    pages = "15",
    year = "2023"
}

@article{Esteban:2024eli,
    author = "Esteban, Ivan and Gonzalez-Garcia, M. C. and Maltoni, Michele and Martinez-Soler, Ivan and Pinheiro, Jo{\~a}o Paulo and Schwetz, Thomas",
    title = "{NuFit-6.0: updated global analysis of three-flavor neutrino oscillations}",
    eprint = "2410.05380",
    archivePrefix = "arXiv",
    primaryClass = "hep-ph",
    doi = "10.1007/JHEP12(2024)216",
    journal = "JHEP",
    volume = "12",
    pages = "216",
    year = "2024"
}

@article{Huang:2019hgs,
    author = "Huang, Guo-yuan and Liu, Qinrui",
    title = "{Hunting the Glashow Resonance with PeV Neutrino Telescopes}",
    eprint = "1912.02976",
    archivePrefix = "arXiv",
    primaryClass = "hep-ph",
    doi = "10.1088/1475-7516/2020/03/005",
    journal = "JCAP",
    volume = "03",
    pages = "005",
    year = "2020"
}

@online{Ding:2026wjp,
    author = "Ding, Tianyi and Liu, Qinrui",
    title = "{Probing Neutrino Flavor Composition with the Glashow Resonance at Tau Air-Shower Neutrino Telescopes}",
    eprint = "2607.26128",
    archivePrefix = "arXiv",
    primaryClass = "hep-ph",
    month = "7",
    year = "2026", 
}

@article{PierreAuger:2019ens,
    author = "Aab, A. and others",
    collaboration = "Pierre Auger",
    title = "{Probing the origin of ultra-high-energy cosmic rays with neutrinos in the EeV energy range using the Pierre Auger Observatory}",
    eprint = "1906.07422",
    archivePrefix = "arXiv",
    primaryClass = "astro-ph.HE",
    reportNumber = "FERMILAB-PUB-19-280-ND-PPD-TD",
    doi = "10.1088/1475-7516/2019/10/022",
    journal = "JCAP",
    volume = "10",
    pages = "022",
    year = "2019"
}

\appendix
\onecolumngrid

\newpage

\section{High, Mid, and Low Flux Scenarios}
\label{app:flux}

\begin{figure}[h]
    \centering
    \includegraphics[width=0.5\linewidth]{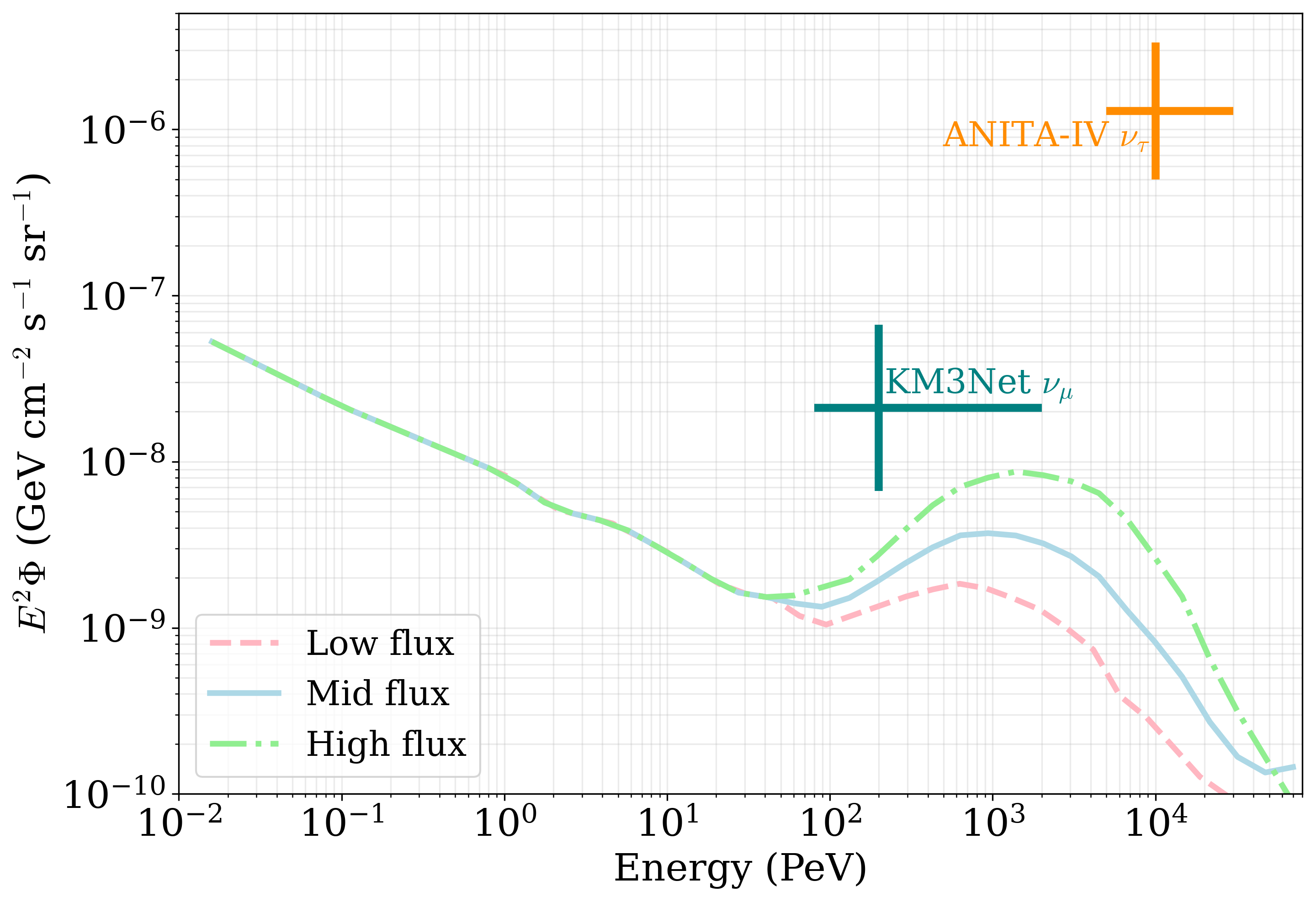}
    \caption{
    Benchmark model of high-energy and ultra-high-energy neutrino flux.
    The model combines the IceCube 9.5-year muon-neutrino measurement with a cosmogenic component. The high-flux scenario uses data from the
    Telescope Array (TA), whereas the low-flux scenario uses the
    cosmogenic UHECR flux inferred from AUGER measurements. \cite{Coleman:2024scd} The mid-flux scenario is defined as the geometric mean of the two models. 
    The teal marker indicates the KM3-230213A event from KM3NeT~\cite{KM3NeT:2025npi},
    consistent with a $\nu_\mu$ origin. The orange markers denote the four
    near-horizon events observed by ANITA-IV~\cite{ANITA:2020gmv},
    consistent with a $\nu_\tau$ origin.
    }
    \label{fig:flux}
\end{figure}

\begin{figure}[!t]
    \centering

    \begin{minipage}{\linewidth}
        \centering
        \includegraphics[width=\linewidth]{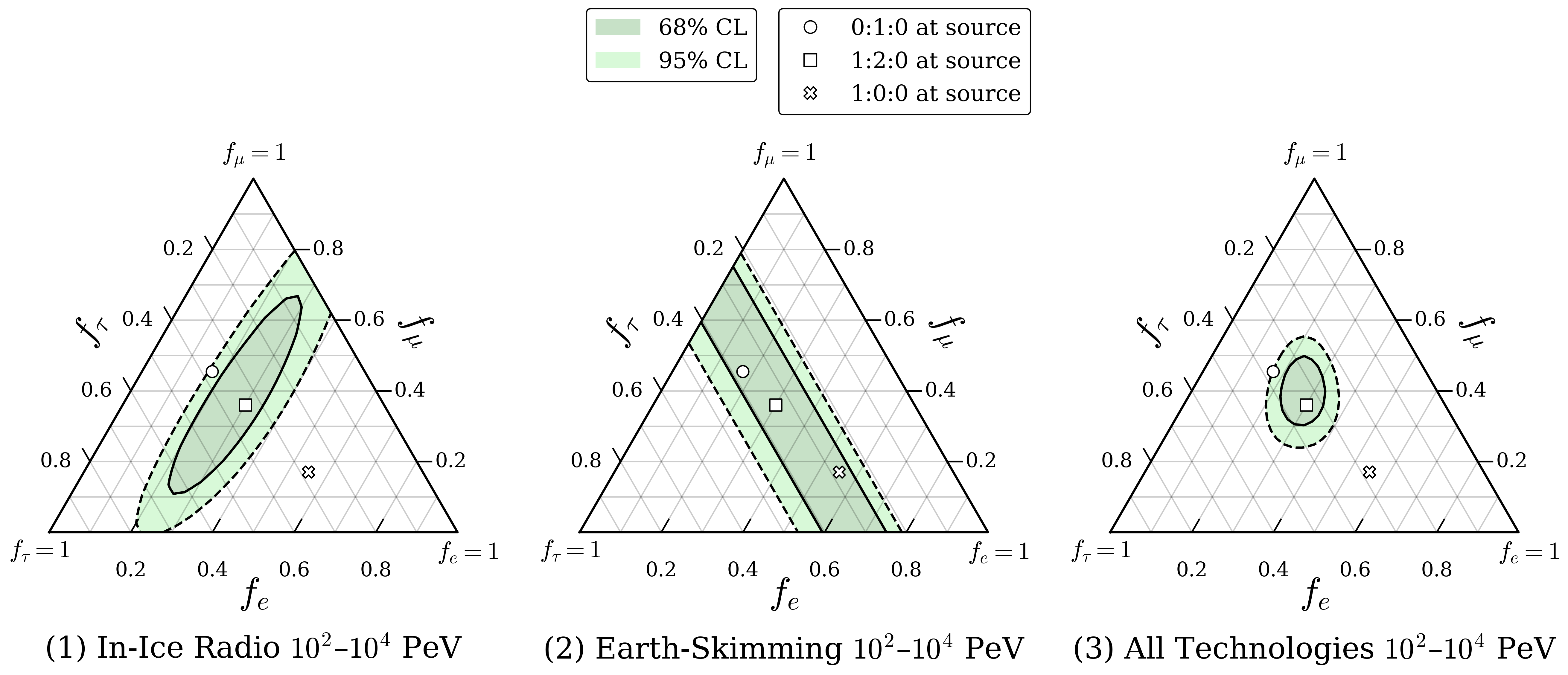}
    \end{minipage}

    \begin{minipage}{\linewidth}
        \centering
        \includegraphics[width=\linewidth]{figure1.png}
    \end{minipage}

    \begin{minipage}{\linewidth}
        \centering
        \includegraphics[width=\linewidth]{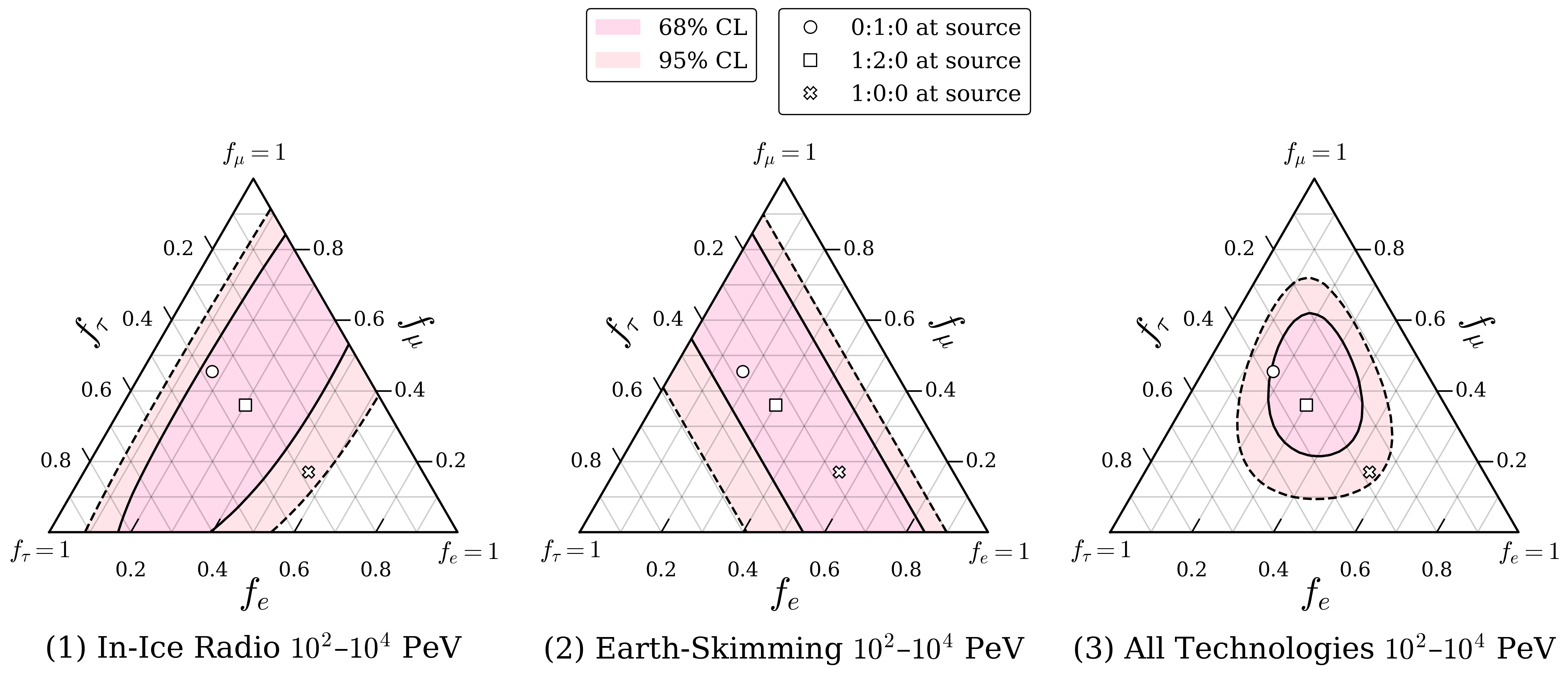}
    \end{minipage}

    \caption{
    Flavor sensitivity to ultra-high-energy neutrinos after 10 years of exposure for three benchmark flux scenarios:
    (a) high-flux, (b) mid-flux, and (c) low-flux. The injected source flavor ratio $(1\!:\!2\!:\!0)_S$ from pion decay evolves to $(0.30\!:\!0.36\!:\!0.34)$ at Earth. 
    }
    \label{fig:flux_scenarios}
\end{figure}

\clearpage
\section{Earth-Skimming Flavor Sensitivity Calculation}
\label{app:skimming}

\begin{figure}[H]
    \centering
    \includegraphics[width=0.8\linewidth]{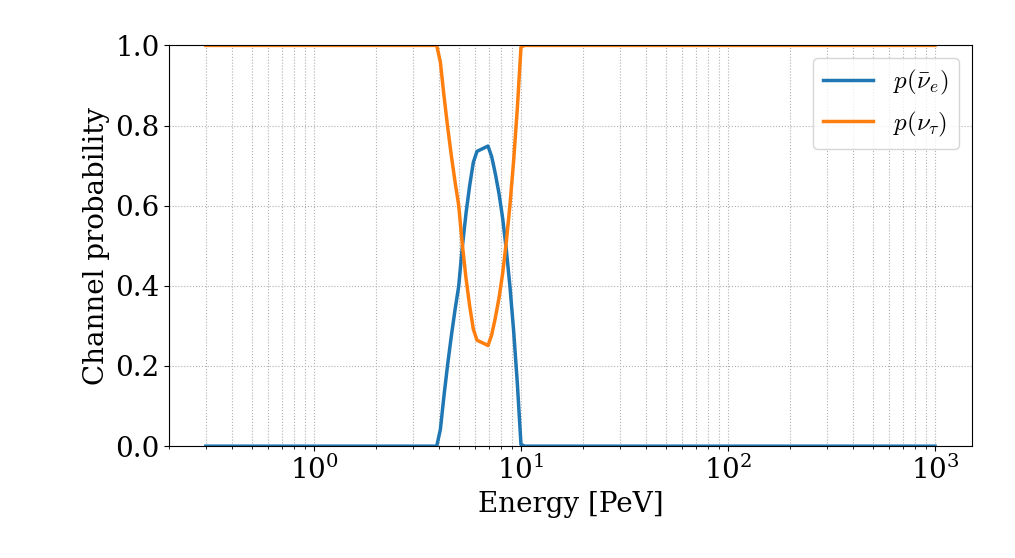}
    \caption{Probability of detecting an event in the $\bar{\nu_e}$ channel relative to the $\nu_\tau$ channel as a function of neutrino energy~\cite{TAMBO:2025jio}.}
    \label{fig:glashow}
\end{figure}

\newpage
\section{In-Ice Radio Flavor Sensitivity Calculation}
\label{app:radio}

The likelihood for the $\nu_e$ CC classification channel is given by  

\begin{equation}
\mathcal{L}_{\text{ML}}(\theta) = 
\prod_{i=1}^{N_E} 
\mathcal{P}\!\left(N_{\nu,\text{all},i}^{\text{obs}}; N_{\nu,\text{all},i}(\theta)\right) 
\, P_{\text{ML},i}(\theta),
\end{equation}

where $\mathcal{P}$ denotes the Poisson distribution for observing $N_{\nu,\text{all},i}^{\text{obs}}$ events, given the expected number $N_{\nu,\text{all},i}(\theta)$. The additional factor  

\begin{equation}
P_{\text{ML},i}(\theta) = \sum_{k=0}^{N_{\text{ML},i}^{\text{obs}}} 
\mathcal{B}\!\left(k; N_{\nu,\text{all},i}^{\text{obs}}, p_{\text{tp},i}(\theta)\right) 
\cdot \mathcal{B}\!\left(N_{\text{ML},i}^{\text{obs}} - k; N_{\nu,\text{all},i}^{\text{obs}}, p_{\text{fp},i}(\theta)\right),
\end{equation}

encodes the probability of correctly identifying true-positive ($\nu_e$ CC) versus false-positive events. Here, $\mathcal{B}$ denotes a binomial probability. The true-positive and false-positive probabilities are 

\begin{equation}
p_{\text{tp},i}(\theta) = \frac{N_{\nu_e,i}^{\text{CC}}(\theta)}{N_{\nu,\text{all},i}(\theta)} \, \mathcal{T}_i,
\qquad
p_{\text{fp},i}(\theta) = \frac{N_{\nu,\text{all},i}(\theta) - N_{\nu_e,i}^{\text{CC}}(\theta)}{N_{\nu,\text{all},i}(\theta)} \, \mathcal{F}_i ,
\end{equation}

where $\mathcal{T}_i$ and $\mathcal{F}_i$ represent classification and mis-classification efficiency. See Figure~\ref{fig:fp_rates}, left.

By analogy, the multi-shower channel probability term is

\begin{equation}
P_{\text{mult},i}(\theta) = \sum_{k=0}^{N_{\text{mult},i}^{\text{obs}}}
\mathcal{B}\!\left(k; N_{\nu,\text{all},i}^{\text{obs}}, p_{\mu,i}(\theta)\right)
\cdot \mathcal{B}\!\left(N_{\text{mult},i}^{\text{obs}} - k; N_{\nu,\text{all},i}^{\text{obs}}, p_{\tau,i}(\theta)\right),
\end{equation}

where the probabilities for $\nu_\mu$ and $\nu_\tau$ CC interactions to trigger multi-shower events can be expressed as:   

\begin{equation}
p_{\mu,i}(\theta) = \frac{N_{\nu_\mu,i}^{\text{CC}}(\theta)}{N_{\nu,\text{all},i}(\theta)} \, r_{\mu,i},
\qquad
p_{\tau,i}(\theta) = \frac{N_{\nu_\tau,i}^{\text{CC}}(\theta)}{N_{\nu,\text{all},i}(\theta)} \, r_{\tau,i}.
\end{equation}

Here $r_{\mu,i}$ and $r_{\tau,i}$ are the respective efficiencies for multi-shower identification. See Figure~\ref{fig:fp_rates}, right.

\begin{center}
    \includegraphics[width=\linewidth]{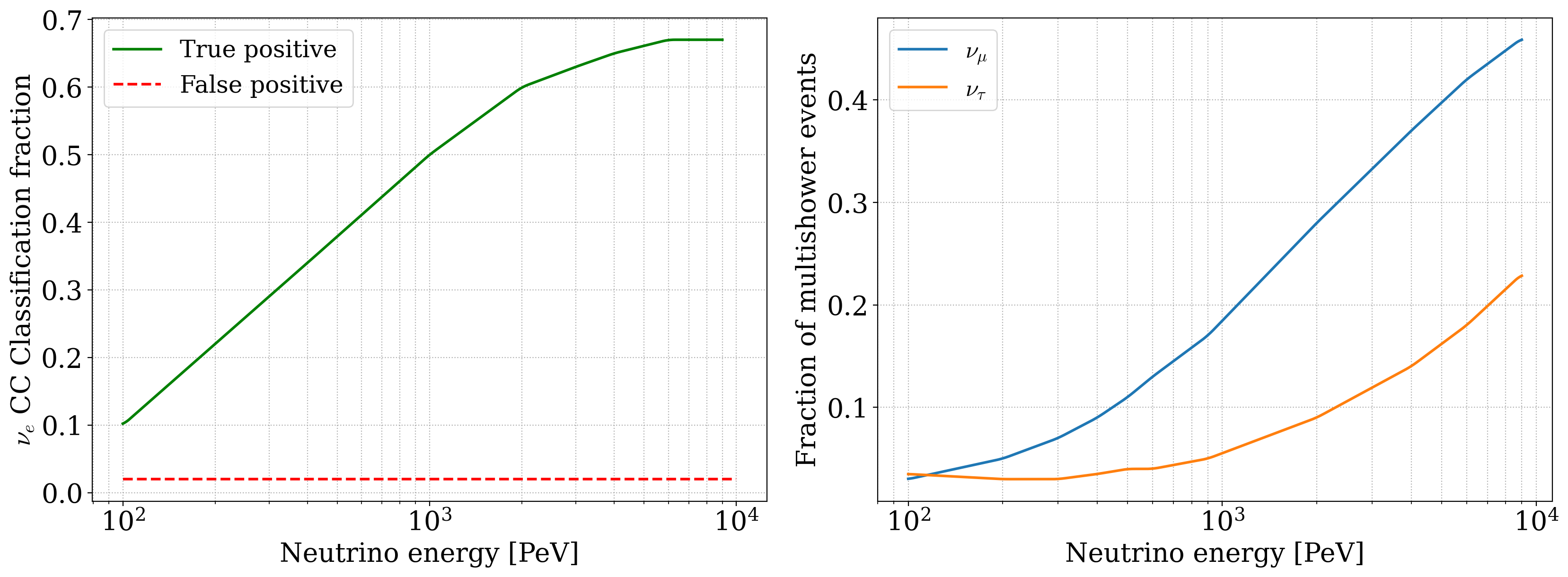}
    \captionof{figure}{Left: True and false positive rates of the convolutional neural network used to classify UHE $\nu_e$ charged-current (CC) events. The network threshold is chosen to yield a 2\% false-positive rate.
Right: Fraction of UHE $\nu_\mu$- and $\nu_\tau$-initiated CC interactions identified as multi-shower events, assuming radio detection in the IceCube-Gen2 array with 2\,km station spacing~\cite{Coleman:2024scd}.}
    \label{fig:fp_rates}
    \end{center}

\begin{figure}[H]
    \centering
    \includegraphics[width=0.6\linewidth]{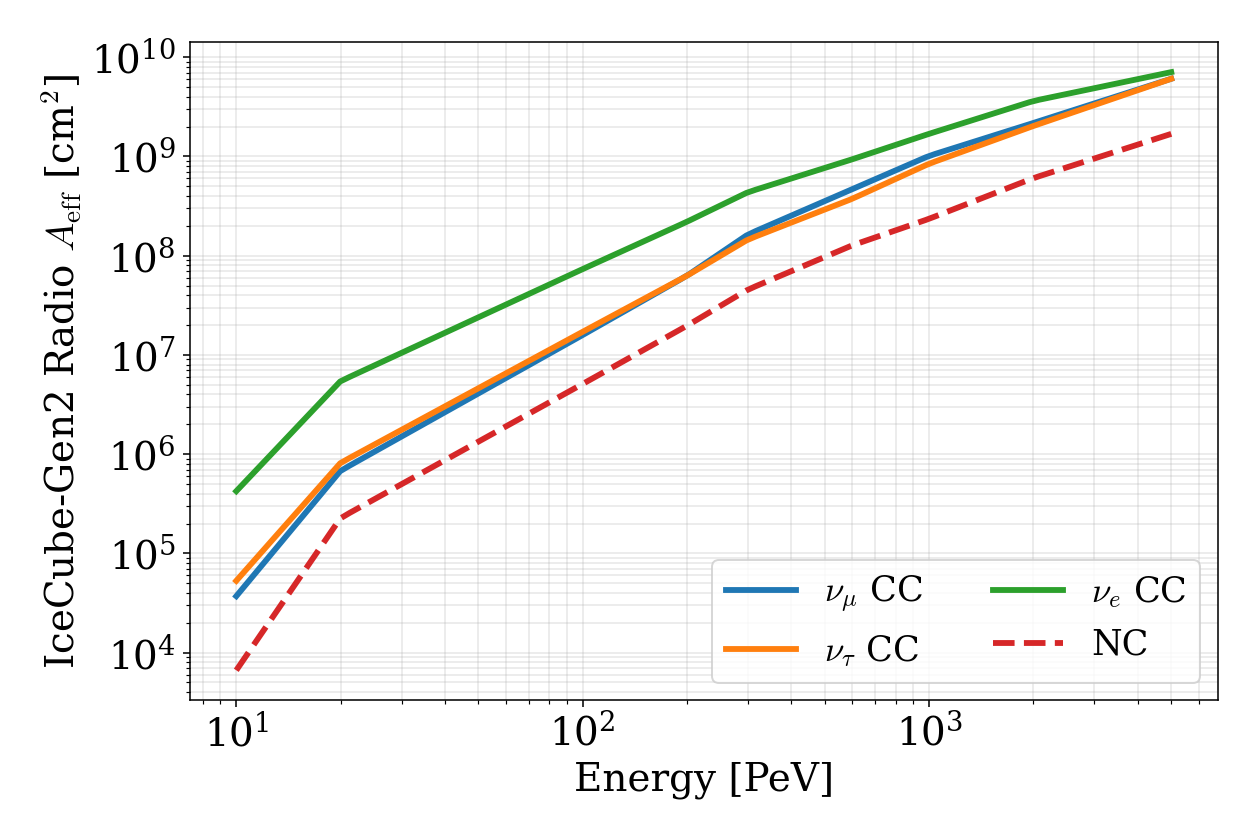}
    \caption{Effective areas of IceCube-Gen2 radio array for ultra-high-energy neutrino detection between 10 and $10^4$ PeV. Areas are shown in $\mathrm{cm}^2$ for charged-current (CC) interactions of $\nu_e$, $\nu_\mu$, and $\nu_\tau$ individually, and for neutral-current (NC) interactions summed over all flavors.}
    \label{fig:RadioEff}
\end{figure}

\end{document}